\newlength{\absize}
\documentclass[12pt]{article}
\usepackage{graphicx}
\usepackage{latexsym}
\usepackage{amssymb}
\usepackage{bbm}
\usepackage{setspace}
\usepackage{braket}
\newcommand{\figsize}{\small}

\input prepictex
\input pictex
\input postpictex
\newdimen\tdim
\tdim=\unitlength
\def\stpltsmbl{\setplotsymbol ({\small .})}

\newbox\sru
\setbox\sru=\hbox{\beginpicture
\setcoordinatesystem units <\tdim,\tdim>
\stpltsmbl
\setquadratic
\plot
  0.0   0.0
  4.8   1.5
  7.5   5.0
  7.3   8.5
  5.0  10.0
  2.7   8.5
  2.5   5.0
  5.2   1.5
 10.0   0.0
/
\endpicture}
\def\springru #1 #2 *#3 /{\multiput {\copy\sru}  at
#1 #2 *#3 10 0 /}

\renewcommand{\bar}{\overline}

\newcommand{\spur}[1]{\!\not\! #1 \,}

\newcommand{\cA}{\mathcal{A}}
\newcommand{\cV}{\mathcal{V}}

\newcommand{\cL}{\mathcal{L}}

\newcommand{\cB}{\mathcal{B}}
\newcommand{\cC}{\mathcal{C}}

\newcommand{\pd}{\partial}

\renewcommand{\slash}[1]{#1\!\!\!/}

\newcommand{\be}{\begin{equation}}
\newcommand{\ee}{\end{equation}}
\newcommand{\bea}{\begin{eqnarray}}
\newcommand{\eea}{\end{eqnarray}}
\newcommand{\nn}{\nonumber}

\newcommand{\comment}[1]{}

\usepackage[hypertex]{hyperref}

\begin{document}

\thispagestyle{empty}
\pagestyle{empty}
\newcommand{\starttext}{\newpage\normalsize
 \pagestyle{plain}
 \setlength{\baselineskip}{3ex}\par
 \setcounter{footnote}{0}
 \renewcommand{\thefootnote}{\arabic{footnote}}
 }
\newcommand{\preprint}[1]{\begin{flushright}
 \setlength{\baselineskip}{3ex}#1\end{flushright}}
\renewcommand{\title}[1]{\begin{center}\LARGE
 #1\end{center}\par}
\renewcommand{\author}[1]{\vspace{2ex}{\large\begin{center}
 \setlength{\baselineskip}{3ex}#1\par\end{center}}}
\renewcommand{\thanks}[1]{\footnote{#1}}
\renewcommand{\abstract}[1]{\vspace{2ex}\normalsize\begin{center}
 \centerline{\bf Abstract}\par\vspace{2ex}\parbox{\absize}{#1
 \setlength{\baselineskip}{2.5ex}\par}
 \end{center}}

%\preprint{}
\title{The Schwinger Point}
\author{
 Howard~Georgi\thanks{\noindent \tt hgeorgi@fas.harvard.edu}
\\ \medskip
Center for the Fundamental Laws of Nature\\
Jefferson Physical Laboratory \\
Harvard University \\
Cambridge, MA 02138
 }
\date{\today}
\abstract{The Sommerfield model with a massive vector field coupled to a
massless fermion in 1+1 dimensions is an exactly solvable analog of a
Bank-Zaks model.  The ``physics'' of the model comprises a massive boson
and an unparticle
sector that survives at low energy as a conformal field theory (Thirring
model).  I discuss the ``Schwinger point'' of the Sommerfield model in
which the vector boson mass goes to zero.  The limit is singular but gauge
invariant quantities should be well-defined.  I give a number of examples,
both (trivially) with local operators and with nonlocal products connected
by Wilson lines (the primary technical accomplishment in this note is the
explicit and very pedestrian calculation of correlators involving straight
Wilson lines).  I
hope that this may give some insight into the nature of bosonization in the
Schwinger model and its connection with unparticle physics which in this
simple case may be thought of as ``incomplete bosonization.''
}

%\newpage
%\tableofcontents
\starttext

\section{Introduction\label{sec-model}}

In~\cite{Georgi:2009xq}, with Kats, we 
explored
techniques for studying the effects of
self-interactions in the conformal sector of an unparticle model. There,
physics is encoded in the higher $n$-point functions of the conformal
theory. We studied inclusive processes and argued that the inclusive
production of unparticle stuff in standard model processes due to the
unparticle self-interactions can be decomposed using the conformal partial
wave expansion and its generalizations into a sum over contributions from
the production of various kinds of unparticle stuff, corresponding to
different primary conformal operators. Such processes typically involve the
production of unparticle stuff associated with operators other than those
to which the standard model couples directly. Thus just as interactions
between particles allow scattering processes to produce new particles in
the final state, so unparticle self-interactions cause the production of
various kinds of unparticle stuff. 
The resulting picture, we believe,
was a step towards understanding what unparticle stuff ``looks like''
because it is somewhat analogous to the way we describe the production and
scattering of ordinary particles in quantum field theory, with the primary
conformal operators playing the role of particles and the coefficients in
the conformal partial wave expansion (and its generalization to include
more fields) playing the role of amplitudes. We illustrated our methods in
the 2D Sommerfield 
model~\cite{Sommerfield:1964,Brown:1963,Thirring-Wess:1964,Dubin-Tarski:1967,Hagen:1967}
that we discussed previously~\cite{Georgi:2008pq} in which the Banks-Zaks 
theory is exactly solvable.

We also discussed explicitly how unparticle interactions at low
energies evolve as the energy increases and showed in detail how the
underlying physics of the Banks-Zaks model appears at high energy.  The
unparticle physics is always there, but as the energy increases, more and
more massive states in the Banks-Zaks model are produced, mocking up the
conventional scaling.

In this modest note, I continue with the study of the Sommerfield model,
and make more explicit the connection with the Schwinger model in the limit
that the vector boson mass in the Lagrangian goes to zero.  
In particular,
I do explicit calculations of correlators involving
straight Wilson lines.  These are possible using the operator solution of the
Sommerfield model.   Even if some of the results are familiar, I hope
readers will find that I have a different way of talking about them
that may be stimulating.

\section{Sommerfield and Thirring\label{sec-sandt}}

We will begin with a review of the Sommerfield model to set notation
which will be slightly different from that in Kats.\footnote{Our
conventions, as
in~\cite{Georgi:2008pq}, are:  
$
g^{00}=-g^{11}=1,\,
\epsilon^{01}=-\epsilon^{10}=-\epsilon_{01}=\epsilon_{10}=1
$. From the defining properties $\{\gamma^\mu,\gamma^\nu\} = 2g^{\mu\nu}$ 
and $\gamma^5 = -\frac{1}{2}\epsilon_{\mu\nu}\gamma^\mu\gamma^\nu$, it follows that
$\gamma^\mu\gamma^5=-\epsilon^{\mu\nu}\gamma_\nu$ and
$\gamma^\mu\gamma^\nu=g^{\mu\nu}+\epsilon^{\mu\nu}\gamma^5$, and we will
use the representation 
$
\gamma^0=
\pmatrix{
0&1\cr
1&0\cr
},\;
\gamma^1=
\pmatrix{
0&-1\cr
1&0\cr
},\;
\gamma^5=\gamma^0\gamma^1=
\pmatrix{
1&0\cr
0&-1\cr
}\,$. Then the components $\psi_1$ and $\psi_2$ describe a right-moving and
left-moving fermion, respectively.
Lightcone coordinates are defined by
$$x^\pm={x^0\pm x^1}
\quad\quad
\partial_\pm=\frac{\partial_0\pm\partial_1}{2}$$
$$x^+\partial_++x^-\partial_-=(x^0+x_1)\frac{\partial_0+\partial_1}{2}+(x^0-x_1)\frac{\partial_0-\partial_1}{2}
=x^0\partial_0+x^1\partial_1$$
$$
A^0=\pd^0\cV/m_0-\pd^1\cA/m
\quad\quad
A^1=\pd^1\cV/m_0-\pd^0\cA/m
$$
$$
A_0=\pd_0\cV/m_0+\pd_1\cA/m
\quad\quad
A_1=\pd_1\cV/m_0+\pd_0\cA/m
\quad\quad
A_\pm=\pd_\pm\cV/m_0\pm\pd_\pm\cA/m
$$
} 
The Sommerfield Lagrangian is
\be
\cL_S =
\bar\psi\,(i\spur\pd - e\slash A)\,\psi
-\frac{1}{4}F^{\mu\nu} F_{\mu\nu}
+\frac{m_0^2}{2}A^\mu\, A_\mu
\label{Sommerfield-model2}
\ee
It will be useful for comparison to consider the corresponding Lagrangian
without the $A^\mu$ kinetic energy term.
\be
\cL_T =
\bar\psi\,(i\spur\pd - e\slash A)\,\psi
+\frac{m_0^2}{2}A^\mu\, A_\mu
\label{Thirring-model2}
\ee
In (\ref{Thirring-model2}), $A^\mu$ is an auxiliary field proportional to the
vector current
\begin{equation}
A^\mu=\frac{e}{m_0^2}\,\bar\psi\,\gamma^\mu\,\psi
=\frac{e}{m_0^2}\,j^\mu
\label{A-Thirring}
\end{equation}
So (\ref{Thirring-model2}) is equivalent to the Thirring model
\be
\cL_T =
i\,\bar\psi\,\spur\pd \,\psi-\frac{\lambda}{2}\,j^\mu\,j_\mu
\label{Thirring-model-lambda}
\ee
with
\begin{equation}
\lambda=\frac{e^2}{m_0^2}
\label{lambda}
\end{equation}
To solve these models, we decompose $A^\mu$ as
\be
A^\mu = \pd^\mu \cV/m_0 + \epsilon^{\mu\nu}\pd_\nu\cA/m
\label{A-decomposition}
\ee
where
\begin{equation}
m^2=m_0^2+e^2/\pi
%\label{}
\end{equation}
Then we can write
\begin{equation}
\epsilon_{\mu\nu}\partial^\mu A^\nu=
\epsilon_{\mu\nu}\partial^\mu\epsilon^{\nu\beta}\pd_\beta\cA/m
=\partial_\mu\partial^\mu\cA/m
\quad\quad
\partial_\mu A^\mu=\partial_\mu\partial^\mu\cV/m_0
%\label{}
\end{equation}
and the Sommerfield Lagrangian becomes
\be
%\begin{array}{c}
\displaystyle
\cL_S = i\bar\psi\spur\pd\psi -
e\bar\psi\gamma_\mu\psi\left(\pd^\mu\cV/m_0 +
\epsilon^{\mu\nu}\pd_\nu\cA/m\right)
%\\ \displaystyle
 + \frac{1}{2m^2}\cA\,\Box^2\cA +
\frac{1}{2}\pd_\mu\cV\pd^\mu\cV -\frac{m_0^2}{2m^2}
\pd_\mu\cA\pd^\mu\cA
%\end{array}
\label{Sommerfield-AV}
\ee
while the Thirring Lagrangian is just missing the $\Box^2$ term
\be
%\begin{array}{c}
\displaystyle
\cL_T = i\bar\psi\spur\pd\psi -
e\bar\psi\gamma_\mu\psi\left(\pd^\mu\cV/m_0 +
\epsilon^{\mu\nu}\pd_\nu\cA/m\right)
%\\ \displaystyle
 + 
\frac{1}{2}\pd_\mu\cV\pd^\mu\cV -\frac{m_0^2}{2m^2}
\pd_\mu\cA\pd^\mu\cA
%\end{array}
\label{Thirring-AV}
\ee

If we change the fermionic variable to
\be
\Psi = e^{ie\left(\cV/m_0 + \cA\gamma^5/m\right)}\psi
\label{psi-redef}
\ee
the fermion becomes free:
\be
\cL_S=i\bar\Psi\spur\pd\Psi + \frac{1}{2}\pd_\mu\cV\pd^\mu\cV +
\frac{1}{2m^2}\cA\,\Box^2\cA - \frac{1}{2}\pd_\mu\cA\pd^\mu\cA
\label{Sommerfield-redefined}
\ee
\be
\cL_T=i\bar\Psi\spur\pd\Psi + \frac{1}{2}\pd_\mu\cV\pd^\mu\cV +
 - \frac{1}{2}\pd_\mu\cA\pd^\mu\cA
\label{Thirring-redefined}
\ee
In the last terms in both (\ref{Sommerfield-redefined}) and
(\ref{Thirring-redefined}), 
$m_0^2/m^2$ has been replaced
by 1
in order to account for the fact that the path integral measure is not
invariant under the $\cA$ part of (\ref{psi-redef})~\cite{Roskies:1980jh}.\footnote{The same
effect gives mass $e/\sqrt\pi$ to the gauge boson in the Schwinger
model. See also~\cite{Georgi:1971iu}.}

Focusing on $\cA$ in (\ref{Sommerfield-redefined}), we can replace it with somewhat more normal looking fields
as follows.
\be
\frac{1}{2m^2}\cA\,\Box^2\cA - \frac{1}{2}\pd_\mu\cA\pd^\mu\cA
\to
-\frac{m^2}{2}\cB^2+\cB\Box\cA - \frac{1}{2}\pd_\mu\cA\pd^\mu\cA
%\label{Sommerfield-redefined}
\ee
\be
=-\frac{m^2}{2}\cB^2+\frac{1}{2}\pd_\mu\cB\pd^\mu\cB
- \frac{1}{2}\pd_\mu\cC\pd^\mu\cC
%\label{Sommerfield-redefined}
\ee
where $\cC=\cA+\cB$, so $\cB$ is a massive field
and $\cC$ is a massless ghost.
In the Thirring Lagrangian, $\cA$ is already a ghost, so we can just
replace $\cA\to\cC$ and
the Lagrangians become
\be
\cL_S=i\bar\Psi\spur\pd\Psi + \frac{1}{2}\pd_\mu\cV\pd^\mu\cV 
-\frac{m^2}{2}\cB^2+\frac{1}{2}\pd_\mu\cB\pd^\mu\cB
- \frac{1}{2}\pd_\mu\cC\pd^\mu\cC\\
\label{Sommerfield-redefined2}
\ee
\be
\cL_T=i\bar\Psi\spur\pd\Psi + \frac{1}{2}\pd_\mu\cV\pd^\mu\cV 
- \frac{1}{2}\pd_\mu\cC\pd^\mu\cC\\
\label{Thirring-redefined2}
\ee
and the original fermion and vector fields can be written in terms of free fields
\be
\psi_S = e^{-ie\left(\cV/m_0 + (\cC-\cB)\gamma^5/m\right)}\Psi
 \quad\quad
\psi_T = e^{-ie\left(\cV/m_0 + \cC\gamma^5/m\right)}\Psi
\label{psi-redef-vcb}
\ee
\be
A_S^\mu = \pd^\mu \cV/m_0 + \epsilon^{\mu\nu}\pd_\nu(\cC-\cB)/m
\quad\quad
A_T^\mu = \pd^\mu \cV/m_0 + \epsilon^{\mu\nu}\pd_\nu\cC/m
\label{avbc}
\ee
Thus the Thirring model is just the Sommerfield model without the $\cB$
field!  This makes sense because it is physically obvious that the
Sommerfield model goes to the Thirring model in the limit $m_0\to\infty$
with $e/m_0$ fixed, but (\ref{A-Thirring}), (\ref{psi-redef-vcb}) and
(\ref{avbc}) make the correspondence very explicit.

We can use  (\ref{psi-redef-vcb}) and (\ref{avbc}) straightforwardly to write down the Green's functions of
both models.  This is done in appendix~\ref{sec-g1}.

\section{The Schwinger Point\label{sec-schwinger}}

There a much less trivial limit of the Sommerfield model --- the
limit $m_0\to0$ with $m$ fixed.  The $m_0=0$ theory is the Schwinger
model~\cite{Schwinger:1962tp}, 
invariant under gauge
transformations:
\begin{equation}
\psi\to e^{i\theta}\,\psi
\quad
A^\mu\to A^\mu-\frac{\partial\theta}{e}
%\label{}
\end{equation}
But the limit $m_0\to0$ is potentially singular because the formal
gauge invariance of the $m_0=0$ theory means that there is no physical
degree of freedom associated with the $A^\mu$ field.  This shows up in the
factors of $1/m_0$ in the   
$A^\mu$ propagator.
However, the singular
piece is a pure gauge.  As long as we calculate only gauge invariant
quantities (including appropriate Wilson lines~\cite{Wilson:1974sk}), nothing will depend on
this and the limit should makes sense 
and go over smoothly to corresponding calculations in the Schwinger model~\cite{Schwinger:1962tp}.
We should be able to see that the fermions are confined --- or ``bosonized''~\cite{Coleman:1974bu} --- and
understand how the unparticle sector disappears and a mass gap appears.

The first comment is that to have any hope of constructing a gauge
invariant quantity, we can only look at objects with fermion number zero.
For these,
it is easy to see how this works for the $\cV$ field part of $A^\mu$ where
the contribution from a Wilson line can completely cancel the $\cV$
dependence and get rid of everything that is singular as $m_0\to0$, so
the limit should be well defined.  
Conversely, if the fermion number is not zero, there is no way to cancel
the $\cV$ dependence and this implies that these things will not be
well-defined as $m_0\to0$.

The simplest interesting things to look at are the correlations of the
local ``unparticle'' operators
\begin{equation}
O_{21}(x)\equiv 
\psi_2^*(x)\,\psi_1(x)
\quad\mbox{and}\quad
O_{12}(x)\equiv 
\psi_1^*(x)\,\psi_2(x)
\label{upops}
\end{equation}
These are gauge invariant and should make sense in the Schwinger limit.
First consider the 2pt function,
\begin{equation}
\braket{0|{\rm T}\,O_{12}(x_1)\,O_{21}(x_2)|0}
\label{2pt-local}
\end{equation}
We can read off (\ref{2pt-local}) from
figure~\ref{fig-1} with $C_0$ set equal to 1 and the result is
\begin{equation}
S_1(x_1-x_2)\,S_2(x_1-x_2)\,C(x_1-x_2)^4
\label{local-result1}
\end{equation}
\begin{equation}
=\frac{1}{4\pi^2}
\,\exp\left(\frac{2e^2}{\pi m^2}
\left(K_0\left(m\sqrt{-(x_1-x_2)^2 + i\epsilon}\right) + \ln\left(\xi m\right)\right)\right)
\,
\left(\frac{1}{-(x_1-x_2)^2+i\epsilon}\right)^{1-(e^2/\pi)/m^2}
\label{local-result2}
\end{equation}
where 
\begin{equation}
\xi\equiv e^{\gamma_E}/2
\label{xi}
\end{equation}
At short distances, $C(x_1-x_2)\to1$ in (\ref{local-result1}) and the result
goes to a product of free fermion propagators.
But in (\ref{local-result2}) 
at long distances you can see clearly the magic result of the Schwinger limit of the
Sommerfield model.  When $m^2=e^2/\pi$, the last term in
(\ref{local-result2}) goes to 1 and only the massive propagator survives.
But for $m^2>e^2/\pi$, we see the unparticle contribution at long distances.

The magic at $m^2=e^2/\pi$ is responsible for one of the more confusing
features of the Schwinger point.  If (\ref{local-result2}) is to satisfy
cluster decomposition, the operators
must have non-zero vacuum expectation values, because it must be that
\begin{equation}
\braket{0|{\rm T}\,O_{12}(x_1)\,O_{21}(x_2)|0}
\mathop{\longrightarrow}\limits_{-(x_1-x_2)^2\to\infty}
\braket{0|O_{12}(x_1)|0}\,\braket{0|O_{21}(x_2)|0}
\label{2pt-local-cluster}
\end{equation}
This means the vacuum at the Schwinger point must be degenerate with
\begin{equation}
\braket{0|O_{12}(x_1)|0}
=\frac{\xi m}{2\pi}\,e^{i\theta}
\quad\quad
\braket{0|O_{21}(x_2)|0}
=\frac{\xi m}{2\pi}\,e^{-i\theta}
\label{2pt-local-vev}
\end{equation}
where $\theta$ is the parameter that labels the vacuum
state.~\cite{Smilga:1992hx,Jayewardena:1988td,Hetrick:1988yg} 
Unless something else is coupled to the unparticle
operators, (\ref{upops}), such as a mass term, a source, or a more
complicated interaction, there is no physics in these VEVs.
They must be there for the theory to be consistent with cluster
decomposition, but they have no other consequences.

The tools in the appendix (and \cite{Georgi:2009xq}) can be used to show
that the behavior we see in (\ref{local-result2}) 
persists in correlation functions involving more than
two of the local
unparticle operators, (\ref{upops}).  In the free-field description of
section~\ref{sec-model}, the local unparticle operators are
\begin{equation}
O_{21}(x)=
\psi_2^*(x)\,\psi_1(x)
=\Psi_2^*(x)\,\Psi_1(x)\,e^{-2ie\cA/m}
=\Psi_2^*(x)\,\Psi_1(x)\,e^{-2ie(\cC-\cB)/m}
\label{upops-ff21}
\end{equation}
\begin{equation}
O_{12}(x)\equiv 
\psi_1^*(x)\,\psi_2(x)
=\Psi_1^*(x)\,\Psi_2(x)\,e^{2ie\cA/m}
=\Psi_1^*(x)\,\Psi_2(x)\,e^{2ie(\cC-\cB)/m}
\label{upops-ff12}
%\label{}
\end{equation}
In the Schwinger limit, the $\Psi$ and $\cC$ contributions conspire to give
constant contribution to all long-distance correlators of these objects, so that all the
physics (except the VEVs, (\ref{2pt-local-vev})), is in the exponentials involving the massive field, $\cB$,
\begin{equation}
e^{\pm2ie\cB/m}
\label{expB}
\end{equation}

How does the magic result in the Schwinger model fit in with bosonization?  It seems that we
can create perfectly well-defined operators out of the local fields in
which the massless degrees of freedom show up at short distances. In the local limit, there is
nothing fermionic about it, but the short distance limit of
(\ref{local-result2}) looks like it arises from a pair of massless fermions.  Where does this
come from in a theory with a mass gap?  Clearly, it is a large energy
phenomenon.  The large momentum behavior of the K\"all\'en-Lehman
representation
is obtained asymptotically because the exponentials, (\ref{expB}), produce
more and more massive vector states as the energy increases.\footnote{There
are lots of less trivial examples worked out in \cite{Georgi:2009xq}).}

In more detail, what is happening is that the exponential of the unparticle ghost exactly
compensates for the bi-fermion contribution in (\ref{local-result1}).  At smaller
$e^2$, the compensation is not exact.  The fermion wins and one has an
anomalous dimension for the unparticle operator.  For larger $e^2$, the
ghost wins and  the theory is not unitary. 

Going in the other direction, from the Schwinger model to the Sommerfield
model, this discussion suggests that we might regard the unparticle sector as the
result of ``incomplete bosonization.''  In the Sommerfield model, for $e^2<\pi m^2$, the ghost fields
do not couple strongly enough to completely eliminate the long-distance
physics of the massless fermion fields.  The fermions are not confined into
particle bound states.  But
neither do their propagators have poles like normal particles.  They are
unparticles.  

Although it is not the primary thrust of this paper, it is worth mentioning
what happens to this discussion of the local unparticle operators if we
generalize the Schwinger model to include $n$ massless flavors (see~\cite{Georgiandwarner}).
This model has a classical chiral $U(n)\times U(n)$ symmetry which is presumably
broken by the chiral anomaly down to $SU(n)\times SU(n)\times U(1)$.  At
the Schwinger point, because the vector boson mass gets contributions from
each of the $n$ flavors, $e^2/m^2$ is $1/n$ times what it is in the 1-flavor
Schwinger model.  The ghost contributions to the 
anomalous dimensions of the $(\bar n,n)$ of unparticle
operators (where the first subscript on $\psi$ is the fermion label and the
second subscript indicates the chirality),
\begin{equation}
O^j_{k12}\equiv \psi^*_{j1}\psi_{k2}
\mbox{~~and~~}
O^j_{k21}\equiv \psi^*_{j2}\psi_{k1}
\label{22unparticle}
\end{equation}
are down by $1/n$ compared to what they are in the Schwinger model and so do not cancel the
free fermion contributions to the 2-point functions.  
But the cancellation does take place in the 2-point function of the chiral $SU(n)\times SU(n)$ singlet operators 
\begin{equation}
O^{n-\rm flavor}_{12}\equiv \prod_{\ell=1}^n\psi^*_{\ell1}\psi_{\ell2}
\mbox{~~and~~}
O^{n-\rm flavor}_{21}\equiv \prod_{\ell=1}^n\psi^*_{\ell2}\psi_{\ell1}
\label{n-11unparticle}
\end{equation}
for which
\begin{equation}
%\begin{array}{c}
\displaystyle
\braket{0|{\rm T}\,O^{n-\rm flavor}_{12}(x_1)\,O^{n-\rm flavor}_{21}(x_2)|0}
%\\ \displaystyle 
=\left(\frac{\xi m}{4\pi^2}\right)^{2n}
\,\exp\left(2n\pi
K_0\left(m\sqrt{-(x_1-x_2)^2 + i\epsilon}\right) \right)
%\end{array}
\label{n-2pt-11}
\end{equation}
with $\xi= \frac{e^{\gamma_E}}{2}$ as in (\ref{xi}).
Thus cluster decomposition requires that these operators have
VEVs,\footnote{These issues will be explored in detail in \cite{Georgiandwarner2}.}
\begin{equation}
\braket{0|O^{n-\rm flavor}_{12}(x_1)|0}
=e^{in\theta}\left(\frac{\xi m}{4\pi^2}\right)^{n}
\quad\quad
\braket{0|O^{n-\rm flavor}_{21}(x_2)|0}
=e^{-in\theta}\left(\frac{\xi m}{4\pi^2}\right)^{n}
\label{n-2pt-local-vev}
\end{equation}

\section{Wilson lines\label{sec-wilson}}

If the fermions and antifermions in our operators are separated in
space-time, we need Wilson lines~\cite{Wilson:1974sk} to make things gauge invariant.
Thus, for example, we should be able to look at the VEV
\begin{equation}
\braket{0|{\rm T}\,O_{11}(y,x)|0}
\label{vev}
\end{equation}
\begin{equation}
O_{11}(y,x)\equiv 
{\rm T}\,\psi_1^*(y)\exp\left(-ie\int_x^y\,A_\mu(z)\,dz^\mu\right)\,\psi_1(x)
\label{o11wl}
\end{equation}
I am
particularly interested in space-like Wilson lines because they are the
simplest thing to look at, so (\ref{vev}) could be
all at one time, but if we want to think about anything but a straight path
in 1+1, we need the
time dimension as well, so we do the calculation in general.
Under the gauge transformation this goes to 
\begin{equation}
\braket{0|{\rm
T}\,\psi_1^*(y)\,e^{-i\theta(y)}\,\exp\left(-ie\int_x^y\,A_\mu(z)\,dz^\mu
+i\int_x^y\partial_\mu\theta(z)\,dz^\mu\right)\,e^{i\theta(x)}\,\psi_1(x)|0}
\label{2pt-1}
\end{equation}
\begin{equation}
\braket{0|{\rm
T}\,\psi_1^*(y)\,e^{-i\theta(y)}\,\exp\left(-ie\int_x^y\,A_\mu(z)\,dz^\mu
+i\theta(y)-i\theta(x)\right)\,e^{i\theta(x)}\,\psi_1(x)|0}
\label{2pt2}
\end{equation}
The VEV in (\ref{vev}) is gauge invariant for any path in the
Wilson line, but the value may depend on the path.  For simplicity, we will
calculate it for a straight path from $x$ to $y$, which should give a
Lorentz covariant quantity:
\begin{equation}
z(\alpha)^\mu =(1-\alpha)x^\mu+\alpha y^\mu
\quad
dz^\mu=(y^\mu-x^\mu)\,d\alpha
\label{straight}
\end{equation}
\begin{equation}
z(\alpha)^\mu -x^\mu=\alpha(y^\mu-x^\mu)
\quad
z(\alpha)^\mu -y^\mu=(1-\alpha)(x^\mu-y^\mu)
%\label{}
\end{equation}
and we can use
\begin{equation}
\psi_1(x)=e^{-ie\big(\cV(x)/m_0+\cA(x)/m\big)}\,\Psi_1(x)
%\label{}
\end{equation}
and (\ref{A-decomposition}) to calculate the contribution of the Wilson line.
The general argument above shows that the Wilson line simply cancels the
$1/m_0$ dependence in the anomalous dimension that come from the $\cV$ fields.
So we set these to zero in calculating (\ref{vev}).   
Thus the Wilson line is
\begin{equation}
\left.\exp\left(-ie\int_{z(0)}^{z(1)}\,A_\mu(z(\alpha))\,dz(\alpha)^\mu \right)\right|_{\cV=0}
=\exp\left(-i\frac{e}{m}\int_{z(0)}^{z(1)}\,\epsilon_{\mu\nu}\,\partial_{z(\alpha)}^\nu
\cA\left(z(\alpha)\right)\,dz(\alpha)^\mu \right)
\label{wilsonA}
\end{equation}
To calculate the contribution of the $\cA$ fields to the Wilson line, we
will do the Wick expansion of all the $\cA$ fields in (\ref{vev}).  This is
much easier than it looks for the straight paths, because the $\epsilon^{\mu\nu}$
in (\ref{A-decomposition}) causes many terms to vanish.  For example, all
the terms in which an $\cA$ in the Wilson line is contracted with an $\cA$ in
$\psi$ or $\psi^*$ vanish because all the coordinate dependence is in the
same direction, proportional to $y^\mu-x^\mu$.
For example, if the $\cA\bigl(z(\alpha)\bigr)$ in (\ref{wilsonA}) is
contracted with $\cA(x)$, the result is a function of 
$(z(\alpha)-x)^2$ and the derivative with respect to $z(\alpha)_\nu$ is
porportional to $z(\alpha)^\nu-x^\nu=\alpha(y^\nu-x^\nu)$ which is
orthogonal to $\epsilon_{\mu\nu}\,dz(\alpha)^\mu $. 
Thus for the straight path (\ref{straight}), the VEV (\ref{vev}) is simply the usual contribution to the
2-pt function with the $1/m_0$ terms removed multiplied by the vacuum value
of the Wilson line.

We will now evaluate the Wilson line contribution explicitly.
Lorentz invariance is crucial here and I want to look at space-like Wilson
lines so we will use $F(-x^2)=F(-x^\mu x_\mu)$ for the
$\cA$ 2-pt function which is  
\begin{equation}
F(-x^2)=\frac{1}{2\pi}
\left[K_0\left(m\sqrt{-x^2 + i\epsilon}\right) + \ln\left(e^{\gamma_E} m\sqrt{-x^2 +
i\epsilon}/2\right)\right]
\label{f}
\end{equation}
Now look at the Wick contractions of the Wilson
line, which is 
\begin{equation}
W(x-y)=\exp\left(-\frac{e^2}{m^2}\,Y(x-y)\right)
\label{w}
\end{equation}
where
\begin{equation}
Y(x-y)=\frac{1}{2}\int\,\epsilon_{\mu_1\nu_1}\,dz(\alpha_1)^{\mu_1}
\epsilon_{\mu_2\nu_2}\,dz(\alpha_2)^{\mu_2}
\partial_{z(\alpha_1)}^{\nu_1}
\partial_{z(\alpha_2)}^{\nu_2}F\left(-\left(z(\alpha_1)-z(\alpha_2)\right)^2\right)
\label{y}
\end{equation}
To evade annihilation by the $\epsilon$s, the partial derivatives must both act on the same factor of
$\left(z(\alpha_1)-z(\alpha_2)\right)^2$, so this is
\begin{equation}
=\int\,\epsilon_{\mu_1\nu_1}\,dz(\alpha_1)^{\mu_1}
\epsilon_{\mu_2\nu_2}\,dz(\alpha_2)^{\mu_2}
g^{\nu_1\nu_2}\,
F'\left(-\left(z(\alpha_1)-z(\alpha_2)\right)^2\right)
%\label{}
\end{equation}
\begin{equation}
=-(y-x)^2\,\int\,d\alpha_1\,d\alpha_2\,
F'\left(-(\alpha_1-\alpha_2)^2(y-x)^2\right)
\label{fp1}
\end{equation}
where
\begin{equation}
F'\left(-x^2\right)
=-\frac{1-m \sqrt{-x^2} K_1\left(m
   \sqrt{-x^2}\right)}{4 \pi  x^2}
\label{fp}
\end{equation}
For a space-like Wilson line, (\ref{fp1}) is
\begin{equation}
Y(x-y)=\ell^2\,\int\,d\alpha_1\,d\alpha_2\,
F'\left((\alpha_1-\alpha_2)^2\ell^2\right)
%\label{}
\end{equation}
where $\ell$ is the invariant length, $\sqrt{-(x-y)^2}$.
Now do the $\alpha_2$ integration for fixed $\alpha=\alpha_1-\alpha_2$
\begin{equation}
\alpha_2=\alpha_1-\alpha\,\left\{
\begin{array}{l}
\geq\max\left(-\alpha,0\right)\\
\leq\min\left(1-\alpha,1\right)
\end{array}
\right.
%\label{}
\end{equation}
If $\alpha>0$, this is $[0,1-|\alpha|]$.\\
If $\alpha<0$, this is $[|\alpha|,1]$.\\
So the integral is always $1-|\alpha|$ so it is
\begin{equation}
=\ell^2\,\int_{-1}^1d\alpha\,(1-|\alpha|)\,
F'\left(\alpha^2\ell^2\right)
%\label{}
\end{equation}
\begin{equation}
=2\ell^2\,\int_0^1d\alpha\,(1-\alpha)\,
F'\left(\alpha^2\ell^2\right)
=\int_0^1(1-\alpha)\frac{1-\alpha  l K_1(l \alpha )}{2 \pi  \alpha ^2}\,d\alpha
\label{2pt-result-ell}
\end{equation}
{\figsize\begin{figure}[htb]
$$\includegraphics[width=.7\hsize]{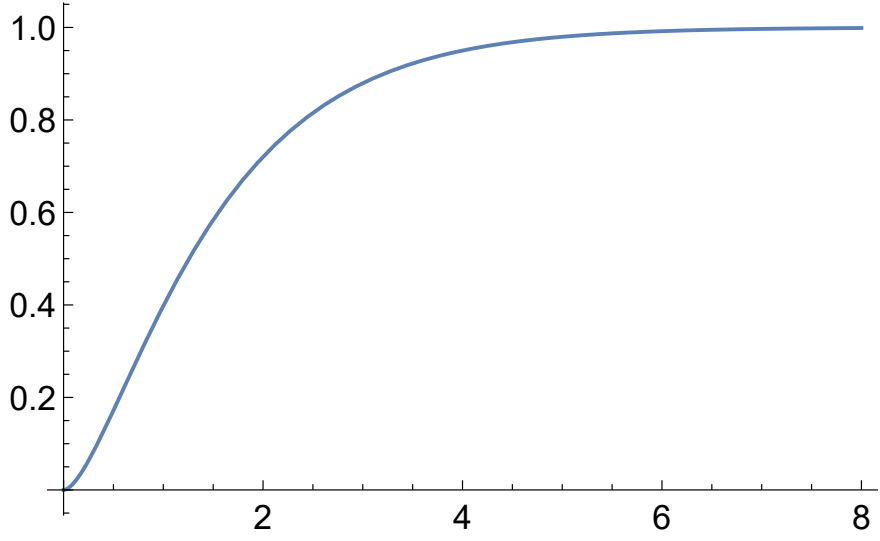}$$
\caption{\figsize\sf\label{fig-fp}$1-\alpha K_1( \alpha )$ versus $\alpha$.}\end{figure}}
It is straightforward to get a qualitative understanding of the large $\ell$
behavior of (\ref{2pt-result-ell}) from the
graph in figure~\ref{fig-fp}. For $\alpha\ll1$, the numerator factor $1-\alpha K_1( \alpha )$
goes to zero because of the cancellation between the ghost and the massive
gauge boson contribution.  For $\alpha\gg1$ it goes to $1$ because the
massive gauge boson contribution vanishes exponentially. The leading
term from the 1 in the $(1-\alpha)$ factor grows linearly with $\ell$ 
because the linear divergence
from the denominator at small  $\alpha$  is cut off for $\alpha\approx1/\ell$.   
The integral can be done explicitly for the $\alpha$ term in $1-\alpha$.  It
grows more negative like a log at long distances because the log divergence
at small $\alpha$ is again cut off for $\alpha\approx1/\ell$.   
Putting the two together and putting the factors of $m$ back 
gives for the large $\ell$ behavior of the integral
\begin{equation}
\frac{m\ell}{4}-\frac{1}{2\pi}\left(\log(e\xi m\ell) \right)+\cdots
%\label{}
\end{equation}
which means that (\ref{vev}) goes to zero exponentially for large $\ell$,
like
\begin{equation}
\begin{array}{c}
\displaystyle
\exp\left(-\frac{e^2\ell}{4m}\right)\,\ell^{e^2/(2\pi m^2)}
\,(e\xi m)^{e^2/(2\pi m^2)}
\\
\displaystyle
=\exp\left(-\frac{\pi(m^2-m_0^2)\ell}{4m}\right)\,\ell^{e^2/(2\pi m^2)}
\,(e\xi m)^{e^2/(2\pi m^2)}
\end{array}
%\label{}
\end{equation} 
so
\begin{equation}
W(x-y)=
\exp\left(-\frac{e^2\sqrt{-(x-y)^2}}{4m}\right)\,\left(-(x-y)^2\right)^{e^2/(4\pi m^2)}
\,(e\xi m)^{e^2/(2\pi m^2)}
\label{w-result}
\end{equation}

The contribution of the Wilson line simply gets multiplied by the usual
contribution from the fermion 2-point function, without the $C_0$ terms.
This can be read off from 
figure~\ref{fig-1} and (\ref{C(x)})-(\ref{s2})
with $C_0$ set equal to 1 and the result is
\begin{equation}
S_1(x-y)\,C(x-y)=
\label{11-result1}
\end{equation}
\begin{equation}
\frac{x^+-y^+}{2\pi}
\,\exp\left(\frac{e^2}{2\pi m^2}
\left(K_0\left(m\sqrt{-(x-y)^2 + i\epsilon}\right) + \ln\left(\xi m\right)\right)\right)
\,
\left(\frac{1}{-(x-y)^2+i\epsilon}\right)^{1-e^2/(4\pi m^2)}
\label{11-result2}
\end{equation}
Putting this all together with $\ell=\sqrt{-(x-y)^2}$
gives 
\begin{equation}
\frac{x^+-y^+}{2\pi}
\,\exp\left(-\frac{e^2\ell}{4m}\right)
\,
\left(\ell\right)^{-2+e^2/(\pi m^2)}\,(e\xi^2 m^2)^{e^2/(2\pi m^2)}
+\cdots
\label{11-result3}
\end{equation}
or
\begin{equation}
\frac{1}{2\pi}\sqrt{-\frac{x^+-y^+}{x^--y^-}}
\,\exp\left(-\frac{e^2\sqrt{-(x-y)^2}}{4m}\right)
\,
\left(-(x-y)^2\right)^{-1/2+e^2/(2\pi m^2)}\,(e\xi^2 m^2)^{e^2/(2\pi m^2)}
+\cdots
\label{11-result3xy}
\end{equation}
where the unwritten terms have additional exponential suppression at large $\ell$.
In the Schwinger limit, $e^2=\pi m^2$, there is only exponential scaling at
long distances.
Thus the Wilson line effectively screens the fermion charges in the
Schwinger limit. 
But as for the correlation functions of the unparticle operators, $O_{12}$
and $O_{21}$, for $e^2<\pi m^2$, there is power-law dependence with
anomalous dimensions at long
distances. 

Now for something more complicated.

\begin{equation}
O_{21}(y,x)\equiv 
{\rm T}\,\psi_2^*(y)\exp\left(-ie\int_x^y\,A_\mu(z)\,dz^\mu\right)\,\psi_1(x)
\label{o21wl}
\end{equation}
\begin{equation}
O_{12}(y,x)\equiv 
{\rm T}\,\psi_1^*(y)\exp\left(-ie\int_x^y\,A_\mu(z)\,dz^\mu\right)\,\psi_2(x)
\label{o12wl}
\end{equation}
\begin{equation}
O_{22}(y,x)\equiv 
{\rm T}\,\psi_2^*(y)\exp\left(-ie\int_x^y\,A_\mu(z)\,dz^\mu\right)\,\psi_2(x)
\label{o22wl}
\end{equation}

The interesting case is the object
\begin{equation}
\braket{0|{\rm T}\,O_{12}(x_1,y_1)\,O_{21}(x_2,y_2)|0}
\label{4pt}
\end{equation}
Now we have two Wilson lines on two straight paths,
\begin{equation}
z_j(\alpha)^\mu =(1-\alpha_j)x_j^\mu+\alpha y_j^\mu
\quad
dz_j^\mu(\alpha_j)=(y_j^\mu-x_j^\mu)\,d\alpha_j
\label{straight2}
\end{equation}
\begin{equation}
z_j(\alpha_j)^\mu -x_j^\mu=\alpha_j(y_j^\mu-x_j^\mu)
\quad
z_j(\alpha_j)^\mu -y_j^\mu=(1-\alpha_j)(x_j^\mu-y_j^\mu)
\quad\mbox{for $j=1$ or $2$}
%\label{}
\end{equation}
The two Wilson lines are the same as before, but now there is a contraction between
the two and between each of them and the $\psi$s on the other operator.

Before tackling this in general, let's look at the
simpler situation in which we keep
$y_2=x_2$.  Now there is only one Wilson line, and to avoid subscripts I will take 
\begin{equation}
x_1=x\quad
y_1=y\quad
x_2=y_2=z
%\label{}
\end{equation}
\begin{equation}
\braket{0|{\rm T}\,O_{12}(y,x)\,O_{21}(z,z)|0}
\label{4pt-semilocal}
\end{equation}

Without the Wilson line we have
\begin{equation}
\Psi_1^*(y)\,e^{i\cA(y)/m}
\Psi_2(x)\,e^{i\cA(x)/m}
\Psi_2^*(z)\,
\Psi_1(z)\,e^{-2i\cA(z)/m}
%\label{}
\end{equation}
which gives
\begin{equation}
C(x-y)^{-1}\,S_1(x-z)\,C(x-z)^2\,S_2(z-y)\,C(y-z)^2
\label{semi-local-result1}
\end{equation}
Now we can define the Wilson line exactly as in (\ref{straight})-(\ref{wilsonA}). 
Again the contractions of an $\cA$ in the Wilson line with $\cA(x)$ or
$\cA(y)$ give no contribution, and the contractions within the Wilson line
give $W(x-y)$, given by (\ref{w-result}), (\ref{y}), and (\ref{2pt-result-ell}).  The
new piece is the contribution of contractions from the Wilson line to
$\cA(z)$, which is $Z(x,y,z)^2$ where
\begin{equation}
Z(x,y,z)=\exp\left(\frac{e^2}{m^2}\,X(x,y,z)\right)
%\label{}
\end{equation}
with $F$ given by (\ref{f}) as usual this is
\begin{equation}
X(x,y,z)=
\int\,\epsilon_{\mu\nu}\,dz(\alpha)^{\mu}
\partial_{z(\alpha)}^{\nu}
F\left(-\left(z(\alpha)-z\right)^2\right)
\label{X}
\end{equation}
\begin{equation}
=
2\,\epsilon_{\mu\nu}\,(y^\mu-x^\mu)\,
(z^\nu-x^\nu)\,\int\,d\alpha\,
F'\left(-\left(z(\alpha)-z\right)^2\right)
\label{x}
\end{equation}
\begin{equation}
=
\frac{1}{2\pi}\,\epsilon_{\mu\nu}\,(y^\mu-x^\mu)\,
(z^\nu-x^\nu)\,\int\,d\alpha\,
\frac{1-m\sqrt{-\left(z(\alpha)-z\right)^2}\,K_1\left(-m\left(z(\alpha)-z\right)^2\right)}{\left(z(\alpha)-z\right)^2}
\label{x2}
\end{equation}
The denominator of (\ref{x2}) is
\begin{equation}
\left(z(\alpha)-z\right)^2=(1-\alpha)^2(x-z)^2+\alpha^2(y-x)^2
-\alpha(1-\alpha)\Bigl((x-y)^2-(x-z)^2-(y-z)^2\Bigr)
%\label{}
\end{equation}
\begin{equation}
=(1-\alpha)(x-z)^2+\alpha(y-z)^2-\alpha(1-\alpha)(x-y)^2
%\label{}
\end{equation}
In this case, if we take all the distances large compared to $1/m$, the
numerator of the integrand goes to 1 and the denominator is integrable and gives
\begin{equation}
\frac
{\log \left(
\frac
{(x-y)^2-(x-z)^2-(y-z)^2+\sqrt{
   \left((x-y)^2\right)^2-2 (x-y)^2
   ((x-z)^2+(y-z)^2)+((x-z)^2-(y-z)^2)^2}
}
{(x-y)^2-(x-z)^2-(y-z)^2-\sqrt{\left((x-y)^2\right)^2-2
   (x-y)^2
   ((x-z)^2+(y-z)^2)+((x-z)^2-(y-z)^2)^2}
}
\right)
   }
{\sqrt{\left((x-y)^2\right)^2-2 (x-y)^2
   ((x-z)^2+(y-z)^2)+((x-z)^2-(y-z)^2)^2}}
\label{xyz-result}
\end{equation}
Note that the square root in the denominator of (\ref{xyz-result}) is the
absolute value of the numerator factor, $\epsilon_{\mu\nu}\,(y^\mu-x^\mu)\,
(z^\nu-x^\nu)$,\footnote{The combination is cyclic.
$
\epsilon_{\mu\nu}\,(y^\mu-x^\mu)\,
(z^\nu-x^\nu)
=\epsilon_{\mu\nu}\,(x^\mu-z^\mu)\,
(y^\nu-z^\nu)
=\epsilon_{\mu\nu}\,(z^\mu-y^\mu)\,
(x^\nu-y^\nu)
$
} so for large distances, (\ref{x}) can be written as
\begin{equation}
\frac{1}{2\pi}\log \left(
\frac
{(x-y)^2-(x-z)^2-(y-z)^2+\epsilon_{\mu\nu}\,(y^\mu-x^\mu)\,
(z^\nu-x^\nu)
}
{(x-y)^2-(x-z)^2-(y-z)^2-\epsilon_{\mu\nu}\,(y^\mu-x^\mu)\,
(z^\nu-x^\nu)
}
\right)
\label{xyz-result2}
\end{equation}
and therefore for large distance
\begin{equation}
Z(x,y,z)=
\left(
\frac
{(x-y)^2-(x-z)^2-(y-z)^2+\epsilon_{\mu\nu}\,(y^\mu-x^\mu)\,
(z^\nu-x^\nu)
}
{(x-y)^2-(x-z)^2-(y-z)^2-\epsilon_{\mu\nu}\,(y^\mu-x^\mu)\,
(z^\nu-x^\nu)
}
\right)^{e^2/(2\pi m^2)}
\label{xyz-result3}
\end{equation}
Note that a parity transformation interchanges $Z$ and $Z^{-1}$.
Clearly, while there is some dependence on the directions of the 2-vectors,
the result is constant as we go to long distance for fixed angles.

Thus the long-distance behavior of (\ref{4pt-semilocal}) 
is (\ref{xyz-result3})
times
(\ref{semi-local-result1}) multplied by the
Wilson line, (\ref{w}) ---
\begin{equation}
\propto Z(x,y,z)^2\,W(x-y)\,C(x-y)^{-1}\,S_1(x-z)\,C(x-z)^2\,S_2(z-y)\,C(y-z)^2
\label{semi-local-proportional}
\end{equation}

If we go to the Schwinger point, $m^2=e^2/\pi$,
we can use cluster decomposition as we did in (\ref{2pt-local-cluster}) to
find the VEV of $O_{12(y,x)}$
\begin{equation}
\braket{0|{\rm T}\,O_{12}(y,x)\,O_{21}(z,z)|0}
\mathop{\longrightarrow}\limits_{-(x-z)^2\to\infty\atop (x-y)^2\;\rm fixed}
\braket{0|O_{12}(x,y)|0}\,\braket{0|O_{21}(z)|0}
\label{4pt-semilocal-cluster}
\end{equation}
Comparing (\ref{semi-local-proportional}) with
(\ref{local-result1}) and (\ref{2pt-local-vev}) and noting that 
\begin{equation}
Z(x,y,z)
\mathop{\longrightarrow}\limits_{-(x-z)^2\to\infty\atop (x-y)^2\;\rm fixed}
1
\label{4pt-z-cluster}
\end{equation}
we see that
\begin{equation}
\braket{0|O_{12}(x,y)|0}
=W(x-y)\,C(x-y)^{-1}\,\frac{\xi m}{2\pi}\,e^{i\theta}
\label{semilocal-vev}
\end{equation}
As $x\to y$, this goes to (\ref{2pt-local-vev}) (as it must) and 
for large distances, this is
\begin{equation}
\sqrt{e}\,\exp\left(-\frac{e^2}{4m}\sqrt{-(x-y)^2}\right)\,\frac{\xi m}{2\pi}\,e^{i\theta}
%\label{}
\end{equation}
which at the Schwinger point goes to
\begin{equation}
\sqrt{e}\,\exp\left(-\pi m\sqrt{-(x-y)^2}/4\right)\,\frac{\xi m}{2\pi}\,e^{i\theta}
%\label{}
\end{equation}

Now back to the fully non-local situation, (\ref{4pt}).  
The contribution without the Wilson lines is
\begin{equation}
C(x_1-y_1)^{-1}\,C(x_2-y_2)^{-1}\,C(x_1-x_2)
\, C(y_1-y_2)\,S_2(x_1-y_2)\,C(x_1-y_2)\,S_1(x_2-y_1)\,C(x_2-y_1)
\label{4pt-bare}
\end{equation}
Now we have two
Wilson lines, which give a factor of
\begin{equation}
W(x_1-y_1)\,W(x_2-y_2)
\label{4pt-wilson}
\end{equation}
There are four contractions in which an $\cA$ in one of the Wilson lines
gets contracted with an $\cA$ associated with one of the fermions in the
other operator.  This is the calculation we just did, so there are two $Z$s and two $Z^{-1}$,
\begin{equation}
Z(x_1,y_1,x_2)\,Z(x_1,y_1,y_2)\,Z(x_2,y_2,x_1)^{-1}\,Z(x_2,y_2,y_1)^{-1}\,
\label{4pt-z}
\end{equation}

The new
piece is the contraction of an $\cA$ in the Wilson line from $x_1$ to $y_1$ with an $\cA$ in the
Wilson line from $x_2$ to $y_2$.  This gives a contribution that looks
familiar in terms of the function $F$ of (\ref{f}):
\begin{equation}
H(x_1,y_1;x_2,y_2)=\exp\left(-\frac{e^2}{m^2}\,Y_{12}\right)
\label{w12}
\end{equation}
\begin{equation}
Y_{12}=\int\,\epsilon_{\mu_1\nu_1}\,dz_1(\alpha_1)^{\mu_1}
\epsilon_{\mu_2\nu_2}\,dz_2(\alpha_2)^{\mu_2}
\partial_{z_1(\alpha_1)}^{\nu_1}
\partial_{z_2(\alpha_2)}^{\nu_2}F\left(-\left(z_1(\alpha_1)-z_2(\alpha_2)\right)^2\right)
\label{y12}
\end{equation}
where
\begin{equation}
\left(z_1(\alpha_1)-z_2(\alpha_2)\right)^{\mu}
=\Bigl((x_1-x_2)+\alpha_1(y_1-x_1)-\alpha_2(y_2-x_2)\Bigr)^\mu
\label{za1a2}
\end{equation}
For simplicity, we will consider the case in which
$z_1(\alpha_1)-z_2(\alpha_2)$ is space-like for all $\alpha_1$ and
$\alpha_2$.  
\begin{equation}
\left(z_1(\alpha_1)-z_2(\alpha_2)\right)^{\mu}
\left(z_1(\alpha_1)-z_2(\alpha_2)\right)_{\mu}
\;\;\mbox{for $0\leq \alpha_1,\alpha_2\leq1$}
\label{spacelike}
\end{equation}

This is a rather restrictive condition in 1+1 dimensions, as
we will see.

\begin{equation}
Y_{12}=2\int\,\epsilon_{\mu_1\nu_1}\,dz_1(\alpha_1)^{\mu_1}
\epsilon_{\mu_2\nu_2}\,dz_2(\alpha_2)^{\mu_2}
\partial_{z_1(\alpha_1)}^{\nu_1}
\left(z_1(\alpha_1)-z_2(\alpha_2)\right)^{\nu_2}
F'\left(-\left(z_1(\alpha_1)-z_2(\alpha_2)\right)^2\right)
\label{y12-2}
\end{equation}
\begin{equation}
\begin{array}{c}
\displaystyle
=2\int\,\epsilon_{\mu_1\nu_1}\,dz_1(\alpha_1)^{\mu_1}
\epsilon_{\mu_2\nu_2}\,dz_2(\alpha_2)^{\mu_2}\,\Biggl(
g^{\nu_1\nu_2}
F'\left(-\left(z_1(\alpha_1)-z_2(\alpha_2)\right)^2\right)
\\
-2
\left(z_1(\alpha_1)-z_2(\alpha_2)\right)^{\nu_1}
\left(z_1(\alpha_1)-z_2(\alpha_2)\right)^{\nu_2}
F''\left(-\left(z_1(\alpha_1)-z_2(\alpha_2)\right)^2\right)\Biggr)
\end{array}
\label{y12-4}
\end{equation}
\begin{equation}
\begin{array}{c}
\displaystyle
=2\int\,\epsilon_{\mu_1\nu_1}\,(y_1-x_1)^{\mu_1}\,d\alpha_1\,
\epsilon_{\mu_2\nu_2}\,(y_2-x_2)^{\mu_2}\,d\alpha_2\,\Biggl(
g^{\nu_1\nu_2}
F'\left(-\left(z_1(\alpha_1)-z_2(\alpha_2)\right)^2\right)
\\
-2
\left(z_1(\alpha_1)-z_2(\alpha_2)\right)^{\nu_1}
\left(z_1(\alpha_1)-z_2(\alpha_2)\right)^{\nu_2}
F''\left(-\left(z_1(\alpha_1)-z_2(\alpha_2)\right)^2\right)\Biggr)
\end{array}
\label{y12-6}
\end{equation}
\begin{equation}
\begin{array}{c}
\displaystyle
=-2\int\,d\alpha_1\,d\alpha_2\,\Biggl(
(y_1-x_1)_\mu(y_2-x_2)^\mu\,
F'\left(-\left(z_1(\alpha_1)-z_2(\alpha_2)\right)^2\right)
\\
-2(g_{\mu_1\mu_2}g_{\nu_1\nu_2}-g_{\mu_1\nu_2}g_{\nu_1\mu_2})\,
(y_1-x_1)^{\mu_1}\,
(y_2-x_2)^{\mu_2}
\\
\left(z_1(\alpha_1)-z_2(\alpha_2)\right)^{\nu_1}
\left(z_1(\alpha_1)-z_2(\alpha_2)\right)^{\nu_2}
F''\left(-\left(z_1(\alpha_1)-z_2(\alpha_2)\right)^2\right)\Biggr)
\end{array}
\label{y12-8}
\end{equation}
Define
\begin{equation}
\begin{array}{c}
L(\alpha_1,\alpha_2)
\equiv -\left(z_1(\alpha_1)-z_2(\alpha_2)\right)^2
\\
=-\Bigl(
(1-\alpha_1)(1-\alpha_2)(x_1-x_2)^2-\alpha_1(1-\alpha_1)(x_1-y_1)^2
-\alpha_2(1-\alpha_2)(x_2-y_2)^2
\\
+\alpha_1\alpha_2(y_1-y_2)^2+\alpha_1(1-\alpha_2)(y_1-x_2)^2
+\alpha_2(1-\alpha_1)(x_1-y_2)^2\Bigr)
\end{array}
%\label{}
\end{equation}

\begin{equation}
\frac{\partial L}{\partial\alpha_1}=-2\frac{\partial z_1(\alpha_1)^\mu}{\partial\alpha_1}
\left(z_1(\alpha_1)-z_2(\alpha_2)\right)_\mu
=-2(y_1-x_1)^\mu\left(z_1(\alpha_1)-z_2(\alpha_2)\right)_\mu
%\label{}
\end{equation}
\begin{equation}
\frac{\partial L}{\partial\alpha_2}=2\frac{\partial z_2(\alpha_2)^\mu}{\partial\alpha_2}
\left(z_1(\alpha_1)-z_2(\alpha_2)\right)_\mu
=2(y_2-x_2)^\mu\left(z_1(\alpha_1)-z_2(\alpha_2)\right)_\mu
%\label{}
\end{equation}
\begin{equation}
\frac{\partial^2 L}{\partial\alpha_1\partial\alpha_2}
=2(y_1-x_1)^\mu(y_2-x_2)_\mu
%\label{}
\end{equation}
\begin{equation}
\frac{\partial^2F}{\partial\alpha_1\partial\alpha_2}
=\frac{\partial}{\partial\alpha_1}\left(\frac{\partial
L}{\partial\alpha_2}\,F'(L)\right)
=\frac{\partial L}{\partial\alpha_1} \,\frac{\partial L}{\partial\alpha_2} \,F''(L)
+2(y_1-x_1)^\mu(y_2-x_2)_\mu\,F'(L)
%\label{}
\end{equation}
Then we can rewrite (\ref{y12-8}) as
\begin{equation}
Y_{12}=
\int\Biggl(-2(y_1-x_1)_\mu(y_2-x_2)^\mu\,\Bigl(F'(L)+2L\,F''(L)\Bigr)
+\frac{\partial L}{\partial\alpha_1} \,\frac{\partial L}{\partial\alpha_2} \,F''(L)
\Biggr)\,d\alpha_1\,d\alpha_2
\label{amazing0}
\end{equation}
\begin{equation}
=
\int\Biggl(-4(y_1-x_1)_\mu(y_2-x_2)^\mu\,\Bigl(F'(L)+L\,F''(L)\Bigr)
+\frac{\partial^2 F}{\partial\alpha_1\,\partial\alpha_2}
\Biggr)\,d\alpha_1\,d\alpha_2
\label{amazing}
\end{equation}

The structure of (\ref{amazing}) is remarkably simple, and the consequences
of this simple form are even simpler and more remarkable.  As long as $L$
is bounded away from zero in the integral (\ref{amazing0}) (which follows
from 
(\ref{spacelike})),
the term
proportional to $(y_1-x_1)_\mu(y_2-x_2)^\mu$ is exponentially suppressed
for distances larger than $1/m$ because the two terms cancel the log term in $F$, leaving
only the Bessel function term.  For the second term, the integral can done
trivially (and, in fact, is independent of the path), and the final result is
\begin{equation}
\begin{array}{c}
\displaystyle
Y_{12}=\Bigl(
F(L(0,0))+
F(L(1,1))-
F(L(1,0))-
F(L(0,1))\Bigr)+\cdots
\\
\displaystyle
=\Bigl(
F(-(x_1-x_2)^2)+
F(-(y_1-y_2)^2)-
F(-(x_1-y_2)^2)-
F(-(y_1-x_2)^2)\Bigr)+\cdots
\end{array}
\label{y12result}
\end{equation}
If all the distances are large and space-like, this is
\begin{equation}
\frac{1}{4\pi}\left(\log\frac{(x_1-x_2)^2(y_1-y_2)^2}{(x_1-y_2)^2(y_1-x_2)^2}\right)
+\cdots
%\label{}
\end{equation}
where the unwritten terms come from the Bessel function and are exponentially supressed
if (\ref{spacelike}) is satisfied
and thus at long distances
\begin{equation}
H(x_1,y_1;x_2,y_2)=\left(\frac{(x_1-y_2)^2(y_1-x_2)^2}{(x_1-x_2)^2(y_1-y_2)^2}\right)^{e^2/(4\pi
m^2)}
\label{4pt-new}
\end{equation}

But it is useful to remember (\ref{y12result}) in its general form, which
gives
\begin{equation}
H(x_1,y_1;x_2,y_2)=\left(
\frac
{e^{F(-(x_1-y_2)^2)}\,
e^{F(-(y_1-x_2)^2)}}
{e^{F(-(x_1-x_2)^2)}\,
e^{F(-(y_1-y_2)^2)}}
\right)^{e^2/
m^2}+\cdots
\label{4pt-general}
\end{equation}
because we can use this form to calculate this contribution even if we put $n$ Wilson
lines together end-to-end.  The $n$ Wilson lines are
\begin{equation}
W(x_j-y_j)=W(z_j-z_{j+1})
%\label{}
\end{equation}
where we have labeled
$x_j=z_j$ , $y_j=z_{j+1}$ for $j=1$ to $n$.  
In addition to the $n$ Wilson lines, we have $n(n$$-$$1)/2$
$H$ factors --- one for each pair of Wilson lines.  
So the result should be
\begin{equation}
\left(\prod_{j=1}^nW(z_j-z_{j+1})\right)\left(\prod_{j<k}H(z_j,z_{j+1};z_k,z_{k+1})\right)
\label{multi-w}
\end{equation}
Naively substituting this into (\ref{4pt-new}) would give
factors of $(-(z_j-z_j)^2)^{e^2/(4\pi m^2)}$.  However, from
(\ref{4pt-general}) we see that these factors should all be replaced by
$1$, because they arise from the exponential of $F(0)=0$.
Interestingly, when this is done, the result is rather simple
\begin{equation}
\exp\Bigl(e^2F(-(z_1-z_{n+1})^2)/m^2\Bigr)
\,\prod_{j=1}^n\left(\frac{W(z_j-z_{j+1})^{e^2/(4\pi m^2}}{\exp\Bigl(e^2F(-(z_j-z_{j+1})^2)/m^2\Bigr)}\right)
\label{multi-w-result}
\end{equation}
If each of the segments is very long compared to $1/m$, this becomes
\begin{equation}
\exp\left(-\frac{e^2}{4m}\sum_{j=1}^n\sqrt{-(z_j-z_{j+1})^2}\right)\,
\left(-(z_1-z_{n+1})^2\right)^{e^2/(4\pi m^2)}\,
(e^n\xi m)^{e^2/(2\pi m^2)}
\label{multi-w-result-long}
\end{equation}
The factors are suggestive when compared to the result for a straight
Wilson line, (\ref{w-result}) . 
The first factor is just the exponential of minus the (now jagged) path length times $e^2/(4m)$,
as in the single Wilson line.  The power-law factor in the middle also appears in
the Wilson line at long distances, (\ref{w-result}).  There are some issues
however.  The last factor of 
(\ref{multi-w-result-long}) differs from the corresponding factor in
(\ref{w-result}) by  
$e^{(n-1)e^2/(2\pi m^2)}$.  This difference
is related, I believe, to failure of the condition (\ref{spacelike}) at the
$n-1$ points where there Wilson lines are joined together. When $y_1=x_2$
in $H(x_1,y_1;x_2,y_2)$, $L$ vanishes in the corner of the integration
region, for $\alpha_1=1$ and $\alpha_2=0$.  Thus we cannot conclude that
the contribution from the first term in (\ref{amazing}) is
exponentially suppressed.  And in fact, if all the Wilson lines are
parallel, it is easy to see analytically that this provides the missing
factors (as it must in this case because we could have calculated the
result for the straight Wilson line by breaking it up in pieces).  
In general, the last factor gets replaced by
\begin{equation}
(e\xi m)\prod_{j=1}^{n-1}\exp\left(\frac{e^2}{2\pi m^2}(1-\theta_j\coth\theta_j)\right)
\label{extra}
\end{equation}
where
\begin{equation}
\theta_j=\mathop{{\rm ArcCosh}}\left(\frac{-(z_{j+2}-z_{j+1})^\mu(z_{j+1}-z_{j})_\mu}
{\sqrt{(z_{j+2}-z_{j+1})^\mu(z_{j+2}-z_{j+1})_\mu\,
(z_{j+1}-z_{j})^\nu(z_{j+1}-z_{j})_\nu}}\right)
\label{thetaj}
\end{equation}
is a measure of the change of direction in 1+1D between the $j$th and
$(j$$+$$1)$st Wilson lines.  Thus one may think of this as some kind of ``curvature
correction.''  The $\theta_j$ dependence cancels the $n$$-$$1$ extra
factors of $e^{e^2/(2\pi m^2)}$ in  (\ref{multi-w-result-long}) when all
the $\theta_j$ vanish, and gives additional suppression for non-zero $\theta_j$.

Finally, one may be tempted to take $z_{n+1}=z_1$ in (\ref{multi-w}) and
create  a gauge invariant Wilson loop.  Unfortunately, in 1+1D, 
this is not consistent with (\ref{spacelike}), which requires that
\begin{equation}
-(z_{j+2}-z_{j+1})^\mu(z_{j+1}-z_{j})_\mu>0\quad\mbox{for all $j=1$ to $n-1$.}
%\label{}
\end{equation}
Thus because a loop in 1+1
requires a
change in the spacial direction and/or time-like Wilson lines, there is always a region
 in the $\alpha$ integration for some of the Wilson lines in which $L$ changes sign, so the result and
the calculation
become complex (in different ways).  The explicit calculation of entire Wilson loops in this model have
been studied in a very different way by Falomir, Gamboa Saravi,  and
Schaposnik in~\cite{Falomir:1981ae}  

\section{Comments\label{sec-comments}}

I hope that focusing on the relationship between the Sommerfield model and
the Schwinger model as I have in this paper may provide a slightly different
approach to some of the fascinating physics of these models.  I hope also that
the simple, explicit calculations done here may find applications in other areas.

\section{Acknowledgements\label{sec-ack}}

I am grateful to Brian Warner for discussions. This work is supported in part by NSF grant
PHY-1719924.

\appendix
\section{Correlation functions\label{sec-g1}}
From \cite{Georgi:2009xq}, we find the non-zero fermion correlators (which
must have equal numbers, $n_1$, of $\psi_1$ and $\psi_1^*$ and
equal numbers, $n_2$, of $\psi_2$ and $\psi_2^*$)
\begin{equation}
\Braket{0|T\,
\left(\prod_{j=1}^{n_1}\psi_1(x_{1j})\,\psi_1(y_{1j})^*\right)\,
\left(\prod_{j=1}^{n_2}\psi_2(x_{2j})\,\psi_2(y_{2j})^*\right)|0}
\label{cfirst}
\end{equation}
\begin{equation}
=
\left(\prod_{j,k}^{n_1}C_0(x_{1j}-y_{1k})\,C(x_{1j}-y_{1k})\,S_1(x_{1j}-y_{1k})\right)
\end{equation}
\begin{equation}
\times
\left(\prod_{j,k}^{n_2}C_0(x_{2j}-y_{2k})\,C(x_{2j}-y_{2k})\,S_2(x_{2j}-y_{2k})\right)
%\label{}
\end{equation}
\begin{equation}
\times
\left(\prod_{j<k}^{n_1}C_0(x_{1j}-x_{1k})^{-1}\,C(x_{1j}-x_{1k})^{-1}\,S_1(x_{1j}-x_{1k})^{-1}\right)
\end{equation}
\begin{equation}
\times
\left(\prod_{j<k}^{n_2}C_0(x_{2j}-x_{2k})^{-1}\,C(x_{2j}-x_{2k})^{-1}\,S_2(x_{2j}-x_{2k})^{-1}\right)
%\label{}
\end{equation}
\begin{equation}
\times
\left(\prod_{j<k}^{n_1}C_0(y_{1j}-y_{1k})^{-1}\,C(y_{1j}-y_{1k})^{-1}\,S_1(y_{1j}-y_{1k})^{-1}\right)
\end{equation}
\begin{equation}
\times
\left(\prod_{j<k}^{n_2}C_0(y_{2j}-y_{2k})^{-1}\,C(y_{2j}-y_{2k})^{-1}\,S_2(y_{2j}-y_{2k})^{-1}\right)
%\label{}
\end{equation}
\begin{equation}
\times
\left(\prod_{j,k}^{n_1,n_2}C_0(x_{1j}-y_{2k})\,C(x_{1j}-y_{2k})^{-1}\right)
\,
\left(\prod_{j,k}^{n_2,n_1}C_0(x_{2j}-y_{1k})\,C(x_{2j}-y_{1k})^{-1}\right)
%\label{}
\end{equation}
\begin{equation}
\times
\left(\prod_{j,k}^{n_1,n_2}C_0(x_{1j}-x_{2k})^{-1}\,C(x_{1j}-x_{2k})\right)
\,
\left(\prod_{j,k}^{n_1,n_2}C_0(y_{1j}-y_{2k})^{-1}\,C(y_{1j}-y_{2k})\right)
\label{clast}
\end{equation}
where for the Sommerfield model
\bea
C_0(x) &=& \exp\left[i\frac{e^2}{m_0^2}\left[D(x) -
D(0)\right]\right]
\propto \left(-x^2+i\epsilon\right)^{-e^2/4\pi m_0^2} \\
C(x) &=& \exp\left[i\frac{e^2}{m^2}\left[(\Delta(x) - \Delta(0)) - (D(x) - D(0))\right]\right] \nn\\
&=&\exp\left[\frac{e^2}{2\pi m^2}
\left[K_0\left(m\sqrt{-x^2 + i\epsilon}\right) + \ln\left(\xi m\sqrt{-x^2 +
i\epsilon}\right)\right]\right]
\label{C(x)}
\eea
with $\xi  = \frac{e^{\gamma_E}}{2}$ as defined in (\ref{xi})
\begin{equation}
S_0^\alpha(x)
=\int \frac{d^2p}{(2\pi)^2}\,e^{-ipx}\,
\frac{p^0-(-1)^\alpha p^1}{p^2+i\epsilon}
=-\frac{1}{2\pi}\frac{x^0-(-1)^\alpha
x^1}{x^2-i\epsilon}
\label{s0}
\end{equation}
\begin{equation}
S_1(x)
=\int \frac{d^2p}{(2\pi)^2}\,e^{-ipx}\,
\frac{p^0+ p^1}{p^2+i\epsilon}
=-\frac{1}{2\pi}\frac{x^0+
x^1}{x^2-i\epsilon}
\label{s1}
\end{equation}
\begin{equation}
S_2(x)
=\int \frac{d^2p}{(2\pi)^2}\,e^{-ipx}\,
\frac{p^0- p^1}{p^2+i\epsilon}
=-\frac{1}{2\pi}\frac{x^0-
x^1}{x^2-i\epsilon}
\label{s2}
\end{equation}
and for the Thirring model, the massive $\Delta$ propagator is absent.
{\figsize\begin{figure}[htb]
$$\includegraphics[width=.7\hsize]{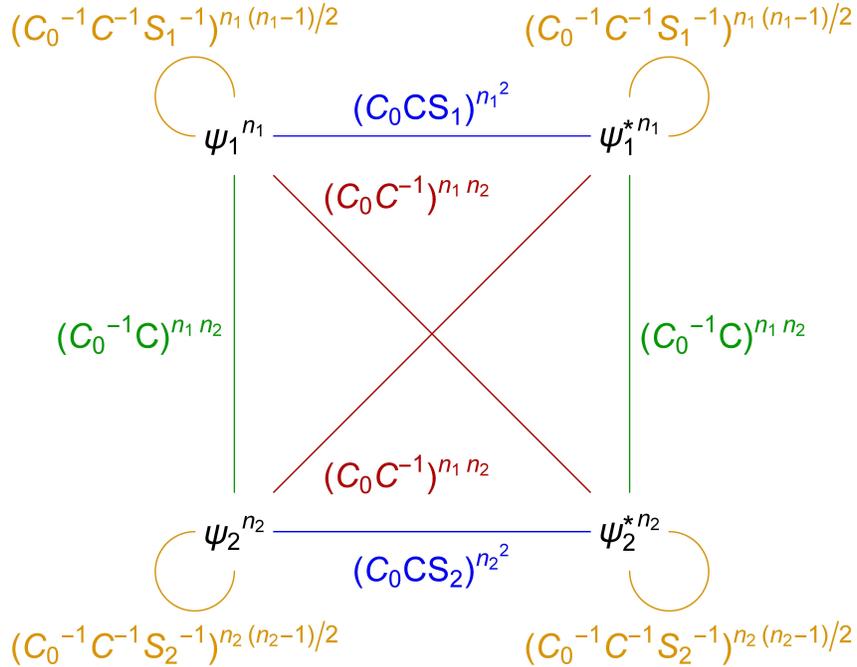}$$
\caption{\figsize\sf\label{fig-1}Pictorial representation of the fermion
correlation functions}\end{figure}}

Many 1+1 miracles go into making this work.  The most miraculous is that we
can write the sum of all the ways of contracting the fermions as a single
term
\begin{equation}
\left(\sum_{P}(-1)^{s(P)}\prod_{j=1}^{n} S_\ell (x_{j}-y_{P(j)})\right)
=(-1)^{n(n-1)/2}\prod_{j,k=1}^{n} S_\ell (x_{j}-y_{k})/\prod_{j<k} S_\ell
(x_{j}-x_{k})/\prod_{j<k} S_\ell (y_{j}-y_{k})
\label{magic}
\end{equation}
for $\ell=1$ or $2$. 
The factors in (\ref{cfirst})-(\ref{clast}) are summarized in the diagram
in figure~\ref{fig-1}.

\bibliography{up4}

%\bibliography{../tex/cft/up4}

\end{document}